\newcommand{\um}{$\mu$m}

\documentclass[10pt,letterpaper]{article}
\usepackage{opex3}

  \makeatletter
  \providecommand\color[2][]{%
    \GenericError{(gnuplot) \space\space\space\@spaces}{%
      Package color not loaded in conjunction with
      terminal option `colourtext'%
    }{See the gnuplot documentation for explanation.%
    }{Either use 'blacktext' in gnuplot or load the package
      color.sty in LaTeX.}%
    \renewcommand\color[2][]{}%
  }%
  \providecommand\includegraphics[2][]{%
    \GenericError{(gnuplot) \space\space\space\@spaces}{%
      Package graphicx or graphics not loaded%
    }{See the gnuplot documentation for explanation.%
    }{The gnuplot epslatex terminal needs graphicx.sty or graphics.sty.}%
    \renewcommand\includegraphics[2][]{}%
  }%
  \providecommand\rotatebox[2]{#2}%
  \@ifundefined{ifGPcolor}{%
    \newif\ifGPcolor
    \GPcolortrue
  }{}%
  \@ifundefined{ifGPblacktext}{%
    \newif\ifGPblacktext
    \GPblacktexttrue
  }{}%
  \let\gplgaddtomacro\g@addto@macro
  \gdef\gplbacktext{}%
  \gdef\gplfronttext{}%
  \makeatother
  \ifGPblacktext
    \def\colorrgb#1{}%
    \def\colorgray#1{}%
  \else
    \ifGPcolor
      \def\colorrgb#1{\color[rgb]{#1}}%
      \def\colorgray#1{\color[gray]{#1}}%
      \expandafter\def\csname LTw\endcsname{\color{white}}%
      \expandafter\def\csname LTb\endcsname{\color{black}}%
      \expandafter\def\csname LTa\endcsname{\color{black}}%
      \expandafter\def\csname LT0\endcsname{\color[rgb]{1,0,0}}%
      \expandafter\def\csname LT1\endcsname{\color[rgb]{0,1,0}}%
      \expandafter\def\csname LT2\endcsname{\color[rgb]{0,0,1}}%
      \expandafter\def\csname LT3\endcsname{\color[rgb]{1,0,1}}%
      \expandafter\def\csname LT4\endcsname{\color[rgb]{0,1,1}}%
      \expandafter\def\csname LT5\endcsname{\color[rgb]{1,1,0}}%
      \expandafter\def\csname LT6\endcsname{\color[rgb]{0,0,0}}%
      \expandafter\def\csname LT7\endcsname{\color[rgb]{1,0.3,0}}%
      \expandafter\def\csname LT8\endcsname{\color[rgb]{0.5,0.5,0.5}}%
    \else
      \def\colorrgb#1{\color{black}}%
      \def\colorgray#1{\color[gray]{#1}}%
      \expandafter\def\csname LTw\endcsname{\color{white}}%
      \expandafter\def\csname LTb\endcsname{\color{black}}%
      \expandafter\def\csname LTa\endcsname{\color{black}}%
      \expandafter\def\csname LT0\endcsname{\color{black}}%
      \expandafter\def\csname LT1\endcsname{\color{black}}%
      \expandafter\def\csname LT2\endcsname{\color{black}}%
      \expandafter\def\csname LT3\endcsname{\color{black}}%
      \expandafter\def\csname LT4\endcsname{\color{black}}%
      \expandafter\def\csname LT5\endcsname{\color{black}}%
      \expandafter\def\csname LT6\endcsname{\color{black}}%
      \expandafter\def\csname LT7\endcsname{\color{black}}%
      \expandafter\def\csname LT8\endcsname{\color{black}}%
    \fi
  \fi

\usepackage{textcomp}
\usepackage{amsmath}
\usepackage{amssymb}
\usepackage{gensymb}

\begin{document}
\title{Optical binding mechanisms: a conceptual model for Gaussian beam traps}

\author{J. M. Taylor and G. D. Love}
\address{Department of Physics, Durham University, South Road, Durham DH1 3LE, United Kingdom}
\email{j.m.taylor@dur.ac.uk}

\begin{abstract}
Optical binding interactions between laser-trapped spherical microparticles are familiar in a wide range of trapping configurations. Recently it has been demonstrated that these experiments can be accurately modeled using Mie scattering or coupled dipole models. This can help confirm the physical phenomena underlying the inter-particle interactions, but does not necessarily develop a conceptual understanding of the effects that can lead to future predictions. Here we interpret results from a Mie scattering model to obtain a physical description which predict the behavior and trends for chains of trapped particles in Gaussian beam traps. In particular, it describes the non-uniform particle spacing and how it changes with the number of particles. We go further than simply \emph{demonstrating} agreement, by showing that the mechanisms ``hidden'' within a mathematically and computationally demanding Mie scattering description can be explained in easily-understood terms.
\end{abstract}



\bibliographystyle{unsrt}

\section{Introduction}

The term optical binding describes the light-mediated lateral or longitudinal interactions between groups of microparticles trapped by one or more laser beams~\cite{burns-binding-science}.  Through their coherent scattering properties, particles will modify the light field surrounding them, and in this way multiple particles are able to interact through the medium of the electromagnetic field around them. This gives rise to a rich tapestry of nonlinear static and dynamic behavior, including chain formation~\cite{Tatarkova, Metzger:fluoro}, bistability~\cite{Metzger:bistability2}, two-dimensional ``crystal'' arrays~\cite{Mellor:polarization} and periodic particle motion~\cite{ng:clusters, cp-dynamics}.

Attempts have been made to model such experiments using Mie scattering models~\cite{ng:clusters, cp-dynamics, us:evanescent} and coupled dipole models~\cite{bessel-binding}. Extremely good agreement was recently reported between a Mie scattering model and the behavior of particles trapped in a counter-propagating Gaussian fiber-based trap~\cite{cp-dynamics}, and a model based on coupled dipole calculations has also recently been to inform a simple explanation of the mechanisms which lead to chain formation in a counter-propagating Bessel beam trap~\cite{bessel-binding}.

In this paper we will use numerical results from a Mie scattering model, as well as from a simpler heuristic model, to build up a detailed understanding of the mechanisms which lead to optical binding of a one-dimensional chain of trapped particles in a counter-propagating Gaussian beam trap. This was possibly the first configuration in which 1D optically bound chains were observed~\cite{Tatarkova}. The inter-particle spacing decreases as the number of particles in the trap increases, and a slight anisotropy is observed in the chain whereby the inter-particle spacing towards the center of the chain is smaller than the spacing at the edges of the chain~\cite[Fig. 5]{cp-dynamics}. Any comprehensive model of the optical interaction must be able to explain these results.

We will show that despite the apparent similarities between the trapped chains in this configuration and the trapped chains in counter-propagating Bessel beams~\cite{bessel-binding}, the binding mechanism is entirely different for a Gaussian beam trap, and is dominated by ``radiation pressure'' effects (the \emph{scattering force}~\cite{nanoparticle-review}). Now that it is clear that Mie scattering models can accurately reproduce experimental results first reported nearly seven years ago, the challenge is to interpret those results in easily-understood terms, and to distill out the key mechanisms ``hidden'' within the complex model in order to develop a conceptual understanding of \emph{why} optical binding and chain formation occurs. We will explain how the trapped particles modify the beam shape to support stable chains of particles with non-uniform inter-particle spacings, whose spacing decreases as the number of particles in the trap increases.

\section{Optical binding concepts, and modeling}

When light is incident on a particle, the particle scatters the light, producing a secondary field which radiates in all directions. An important question to be asked is: what governs the spacing of the particles -- the forward-scattered field or the back-scattered field?
As has been observed by numerous authors~\cite{burns-binding-prl, McGloin:1d-binding, cizmar:bessel, bessel-binding}, the interference of back-scattered light with the incident forward-propagating beam leads to a very large number of nearby trapped configurations, separated by either one or half a wavelength depending on the geometry.
In the Gaussian beam trapping geometry we are considering here, the typical particle spacings are considerably larger than the wavelength of the laser~\cite{Tatarkova} (and indeed the particles may be several wavelengths in diameter). Consequently this binding mechanism, which is most significant for particles up to about half a wavelength in diameter, has little effect on the large-scale behavior of the trapped chains, which is determined largely by the forward-scattered light. 

Because of this important distinction between the effects of forward- and back-scattering, all the models discussed in this paper will explicitly treat the two incoherent counter-propagating beams separately: it is important that effects due to forward-scattering of one beam can be separated from back-scattering effects due to the other beam if the mechanisms are to be understood.
Within a theoretical model, we can not only separate the effects of forward- and back-scattering, but can even ``disable'' one of these effects. The Mie scattering model for $N$ particles can be expressed in the following linear algebra form~\cite{mackowski:translation}:
\begin{eqnarray}
{\bf a}^{(i)} &=& {\bf a}^{(i)}_{ext} + \sum_{j \ne i} {\bf F}_{ji} .{\bf s}^{(j)} \nonumber \\
{\bf s}^{(i)} &= &{\bf T} . {\bf a}^{(i)}
\end{eqnarray}
where ${\bf a}^{(i)}$ represents the net field incident on particle $i$, ${\bf a}^{(i)}_{ext}$ the zero-order incident field on particle $i$ due to the external laser field in the absence of other particles and ${\bf s}^{(i)}$ the scattered field from particle $i$; ${\bf F}_{ji}$ is the translation matrix from a basis centered on particle $j$ to one centered on particle $i$, and {\bf T} is the T-matrix representing the scattering properties of the particle, which is diagonal for spherical particles. This expression simply states that the field incident on a particle $i$ is the (coherent) sum of the laser field and the field scattered by all the other particles.

Using this framework, it is easy to alter the model so that backscatter is not taken into account. If the particles are indexed ``upstream'' to ``downstream'' (i.e. with the first particle closest to the laser source), then we simply modify the sum to read:
\begin{equation}
{\bf a}^{(i)} = {\bf a}^{(i)}_{ext} + \sum_{j < i} {\bf F}_{ji} . {\bf s}^{(j)} \nonumber \\
\end{equation}
Although such a model is un-physical, there can be significant benefits in modeling such a situation: a real-world experiment will be subject to Brownian motion of the trapped particles, which will wash out small local minima in the interactions caused by the back-scattered field. In the absence of Brownian motion (i.e. at absolute zero), a simulation can easily become trapped in such a local minimum. Since it is computationally very expensive to include Brownian motion in a simulation, choosing a model which does not include the back-scattered light is a useful compromise in many cases (though the results of these un-physical simulations must be verified using a full physically-realistic model). Other details of the model used can be found in~\cite{us:evanescent}.

\begin{figure}
\centerline{
  \begingroup
  \setlength{\unitlength}{0.0500bp}%
  \begin{picture}(6480.00,4032.00)%
    \gdef\gplbacktext{}%
    \gdef\gplfronttext{}%
    \gplgaddtomacro\gplbacktext{%
      \csname LTb\endcsname%
      \put(1122,660){\makebox(0,0)[r]{\strut{}-0.04}}%
      \put(1122,1104){\makebox(0,0)[r]{\strut{} 0}}%
      \put(1122,1548){\makebox(0,0)[r]{\strut{} 0.04}}%
      \put(1122,1992){\makebox(0,0)[r]{\strut{} 0.08}}%
      \put(1122,2436){\makebox(0,0)[r]{\strut{} 0.12}}%
      \put(1122,2880){\makebox(0,0)[r]{\strut{} 0.16}}%
      \put(1254,440){\makebox(0,0){\strut{} 0}}%
      \put(2063,440){\makebox(0,0){\strut{} 2}}%
      \put(2871,440){\makebox(0,0){\strut{} 4}}%
      \put(3680,440){\makebox(0,0){\strut{} 6}}%
      \put(4489,440){\makebox(0,0){\strut{} 8}}%
      \put(5297,440){\makebox(0,0){\strut{} 10}}%
      \put(6106,440){\makebox(0,0){\strut{} 12}}%
      \csname LTb\endcsname%
      \put(220,2214){\rotatebox{90}{\makebox(0,0){\strut{}Force (pN)}}}%
      \put(3680,110){\makebox(0,0){\strut{}Spacing ($\mu$m)}}%
      \put(3680,3658){\makebox(0,0){\strut{}}}%
    }%
    \gplgaddtomacro\gplfronttext{%
      \csname LTb\endcsname%
      \put(5119,3595){\makebox(0,0)[r]{\strut{}(a) Plane wave, r-going beam only}}%
      \csname LTb\endcsname%
      \put(5119,3375){\makebox(0,0)[r]{\strut{}(b) Plane wave, both beams}}%
      \csname LTb\endcsname%
      \put(5119,3155){\makebox(0,0)[r]{\strut{}(c) Gaussian, r-going beam only}}%
      \csname LTb\endcsname%
      \put(5119,2935){\makebox(0,0)[r]{\strut{}(d) Gaussian, both beams}}%
    }%
    \gplbacktext
    \put(0,0){\includegraphics{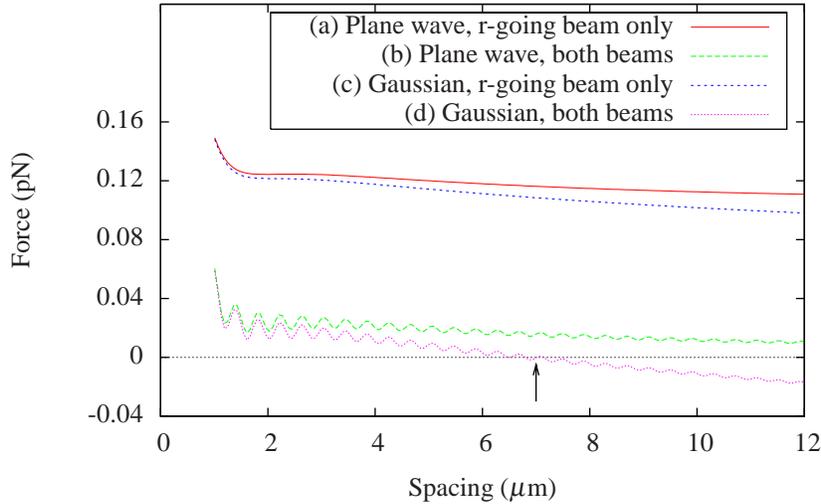}}%
    \gplfronttext
  \end{picture}%
\endgroup
}
\caption{\label{fig:two-particle}Force on a particle in a two particle system, as a function of particle spacing. (a) Force on the downstream (right-hand) one of a pair of 1 \um{} particles illuminated with a single right-going plane wave; (b) repulsive force between the pair of particles when two counter-propagating beams are used; (c) as (a) but for a 25mW Gaussian beam; (d) two counter-propagating Gaussian beams, showing the stable spacing of $\sim 8$ \um{}. Note the modulation due to backscattered light, which has little effect on the general trend of the binding behavior.
The broad harmonic potential introduced by the use of Gaussian beams has altered curve (b) to give curve (d), which is outwardly similar but which has a stable inter-particle spacing at around 7\um{} (marked with an arrow).}
\end{figure}

\section{Chain formation in Gaussian beam traps}

In order to understand the binding mechanisms, consider first the intensity distribution downstream of a single 1.0 \um{} diameter particle illuminated by a single plane wave. The dominant effect is a ``focusing'' of the light (in the limit of large particles we can consider the particle as a spherical lens within the framework of ray optics). Consider now the force on a second particle placed in this field. This is plotted in Figure~\ref{fig:two-particle}, which shows how two particles are stably bound in a Gaussian beam trap, but are not stably bound in counter-propagating plane waves.
We can see that the force due to the light focused by the first particle causes the two particles to be repelled in the case of counter-propagating plane waves (Figure~\ref{fig:two-particle}b).
When we consider counter-propagating Gaussian beams, through symmetry there is no net force on the center of mass of the particle pair, and the  beams provide a broad background harmonic trapping potential. The particles will stabilize with a spacing which is largely determined by the balance of the repulsive force between the two particles and the harmonic trapping potential of the trap (Figure~\ref{fig:two-particle}d), as suggested in~\cite{Tatarkova}. We emphasize that although we refer to a ``focusing'' of the light, we are far from the ray-optics regime, and it is not appropriate to use a ray-optics formula for the focal length, or to suggest that one particle will be bound ``at the focus'' formed by the other particle.

If we just consider the effects of a single beam, then in this experimental setup the contribution of the gradient force turns out to be only a small fraction of the total force on a particle. However, remember that with two counter-propagating beams a large part of the force exerted by one beam is balanced by the force due to the other beam. With some experimental parameters the contribution of the gradient force to the \emph{net force} on a particle due to the two beams together can be non-negligible. Thus it is not really possible to state that either the gradient force or the scattering force will dominate under all circumstances.

\begin{figure}
\centerline{\raisebox{0.5cm}[\depth][\height]{(a)} \hspace{0.7cm} \includegraphics[width=6.7cm]{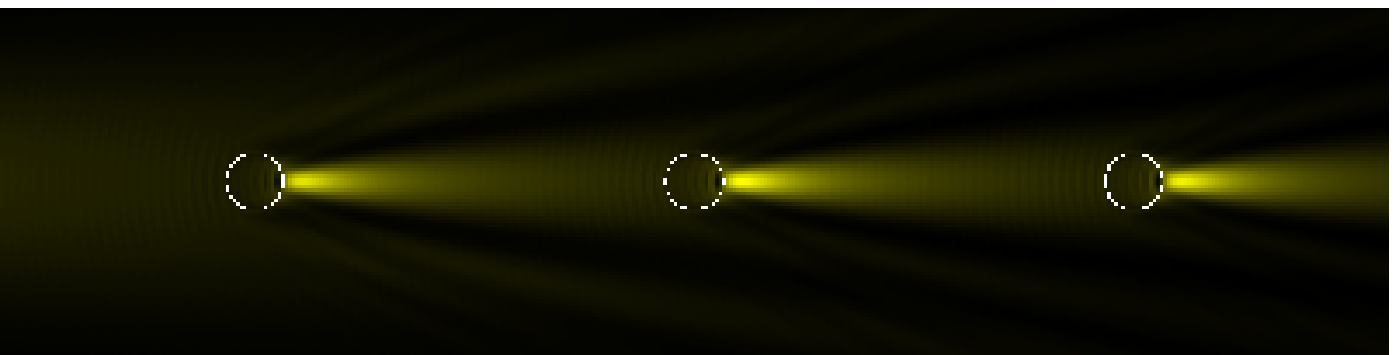}}
\centerline{\raisebox{0.5cm}[\depth][\height]{(b)} \hspace{0.7cm} \includegraphics[width=6.7cm]{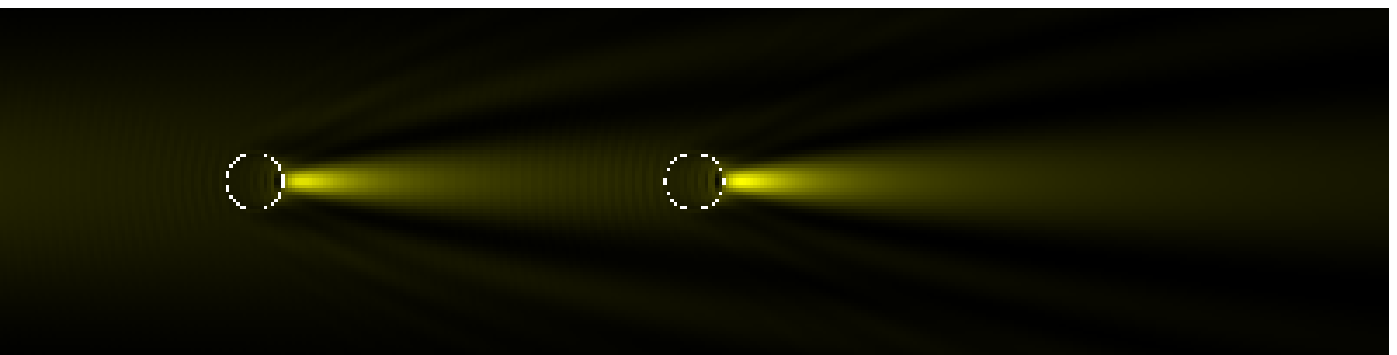}}
\centerline{
  \begingroup
  \setlength{\unitlength}{0.0500bp}%
  \begin{picture}(5760.00,2520.00)%
    \gdef\gplbacktext{}%
    \gdef\gplfronttext{}%
    \gplgaddtomacro\gplbacktext{%
      \csname LTb\endcsname%
      \put(770,660){\makebox(0,0)[r]{\strut{} 0}}%
      \put(770,1458){\makebox(0,0)[r]{\strut{} 0.5}}%
      \put(770,2256){\makebox(0,0)[r]{\strut{} 1}}%
      \put(902,440){\makebox(0,0){\strut{}-30}}%
      \put(2023,440){\makebox(0,0){\strut{}-15}}%
      \put(3144,440){\makebox(0,0){\strut{} 0}}%
      \put(4265,440){\makebox(0,0){\strut{} 15}}%
      \put(5386,440){\makebox(0,0){\strut{} 30}}%
      \put(396,1458){\rotatebox{90}{\makebox(0,0){\strut{}Normalized intensity}}}%
      \put(3144,110){\makebox(0,0){\strut{}Position $z$ ($\mu m$)}}%
      \put(3144,2146){\makebox(0,0){\strut{}}}%
    }%
    \gplgaddtomacro\gplfronttext{%
    }%
    \gplbacktext
    \put(0,0){\includegraphics{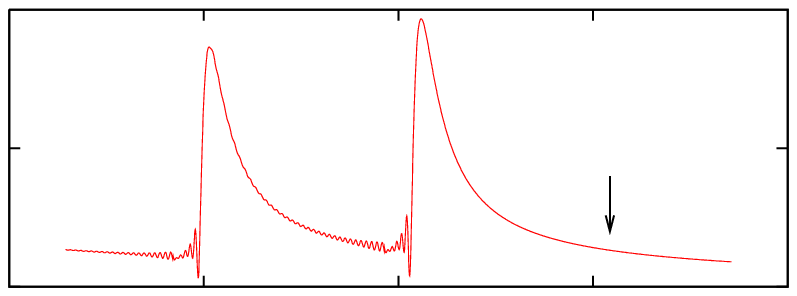}}%
    \gplfronttext
  \end{picture}%
\endgroup
}
\caption{\label{fig:force-from-intensity}A simple Born approximation model uses the intensity of a right-going Gaussian beam in the absence of particle 3 (whose location is indicated with an arrow in (c)) to determine the force $f_3$ on that particle. (a) shows the field for 3 particles exposed to a single Gaussian beam; (b) shows the field in the absence of particle 3 and (c) plots this field (all data in this figure was generated using a full Mie scattering model).}
\end{figure}

A more interesting case than the two-particle case is to consider is that of a larger number $N$ of trapped particles (indexed $i=1$ to $N$), for which we intend to explain the three main properties of the particle chains:
\begin{itemize}
\item The fall in inter-particle spacing with $N$.
\item The anisotropic particle spacing within the chain, with a smaller particle spacing near the middle of the chain.
\item For some experimental parameters, chains are only supported up to a certain number of particles, beyond which the chains collapse.
\end{itemize}
There are a number of statements we can make about the behavior of the chains, based on symmetry considerations, with little or no assumptions on the nature of the inter-particle interactions:
\begin{enumerate}
\item Since the arrangement of the beams (two incoherent beams counter-propagating along the $z$ axis) is symmetric about the $z=0$ plane, the force on particle $i$ due to one beam is equal and opposite to the force on particle $N-i+1$ due to the other counter-propagating beam. This is true for any symmetric arrangement of particles, whether or not this is an equilibrium configuration. If $f_i^+$ (or $f_i^-$) is the force on particle $i$ due to the beam propagating in the $+z$ (or $-z$) direction, then $f_i^+ = f_{N-i+1}^-$.
\item \label{fig:symmetric-profile}In addition, in equilibrium, there must be a net force of zero on each particle when the forces from the two beams are added together (i.e. $f_i^+ = f_i^-$), since by definition there must be no particle motion in equilibrium. Combining this with the previous requirement, we have $f_i^+ = f_{N-i+1}^+$. In other words, the forces on the particles in the chain due to \emph{each individual beam} must be \emph{symmetric} about the center of the chain.
\end{enumerate}
In addition to the above, we will make a number of simplifying hypotheses about the interaction mechanism between the particles. These hypotheses have been observed to be approximately true in numerical experiments:
\begin{enumerate}
\setcounter{enumi}{2}
\item The particle spacings are determined by \emph{forward}-scattering.
\item The force on a given particle $i$ is a function of the light intensity $I_0^{(i)}$ which would be found be at that point \emph{in the absence of} that given particle (the Born approximation; see figure~\ref{fig:force-from-intensity}).
\item \label{item:fixed-profile} The profile of the on-axis scattered intensity downstream of a given particle $i$ has the form $I_0^{(i)} \times I(z-z_i)$, where $I(z-z_i)$ is a \emph{fixed} downstream intensity profile which applies to any particle at any position. Consequently, the force on particle $i+1$ has the form $I_0^{(i)} \times F(z-z_i)$ for a fixed downstream force profile $F(z-z_i)$.
\end{enumerate}
These statements are already enough to explain why the inter-particle spacing at all points in the chain will \emph{decrease} if an additional particle is added onto either end of the chain. The justification for this is as follows. The force pushing inwards on what is now particle 2 in the chain has been increased (it was previously just the force due to the unperturbed laser beam; it is now enhanced by the additional light focused onto it by particle 1). Assuming there are \emph{some} losses along the length of the chain, then the force pushing outwards on what is now the last-but-one particle in the chain will also increase, but by \emph{a smaller amount}. Hence whenever additional particles are added to the chain, the inner ones will be pushed inwards, and so the inter-particle spacing between any given pair of particles in the chain will decrease. Equilibrium is then restored because the closer inter-particle spacing enhances the transmission efficiency, thereby further increasing the repulsive force on the last particle in the chain.

In order to explain the non-uniform particle spacing, we propose a simple ansatz model for the force $f_i$ as a function of $z_i$ and $z_{i-1}$, as follows:
\begin{equation}
f_i(z_i, z_{i-1}) = e^{-\frac{z_i-z_{i-1}}{\alpha}} \times f_{i-1} + (I_0 - \beta z_i)
\end{equation}
Here the first term represents assumption~\ref{item:fixed-profile} and the second term represents a background intensity due to the laser field, which is decreasing with distance from the beam waist. We emphasize that the functional form of $f_i$ has simply been selected empirically  to give a reasonable approximation to the observed inter-particle force. If a closer agreement with the Mie scattering model was desired, a ``hybrid'' model could be used, in which $f_i$ is actually determined from the inter-particle forces for a pair of particles in a plane wave, calculated using Mie scattering theory. However, we have instead chosen to keep our model as elementary as possible.

\begin{figure}
\centerline{
  \begingroup
  \setlength{\unitlength}{0.0500bp}%
  \begin{picture}(5040.00,2016.00)%
    \gdef\gplbacktext{}%
    \gdef\gplfronttext{}%
    \gplgaddtomacro\gplbacktext{%
      \csname LTb\endcsname%
      \put(770,660){\makebox(0,0)[r]{\strut{} 0}}%
      \put(770,1024){\makebox(0,0)[r]{\strut{} 0.5}}%
      \put(770,1388){\makebox(0,0)[r]{\strut{} 1}}%
      \put(770,1752){\makebox(0,0)[r]{\strut{} 1.5}}%
      \put(1320,440){\makebox(0,0){\strut{} 1}}%
      \put(1738,440){\makebox(0,0){\strut{} 2}}%
      \put(2157,440){\makebox(0,0){\strut{} 3}}%
      \put(2575,440){\makebox(0,0){\strut{} 4}}%
      \put(2993,440){\makebox(0,0){\strut{} 5}}%
      \put(3411,440){\makebox(0,0){\strut{} 6}}%
      \put(3830,440){\makebox(0,0){\strut{} 7}}%
      \put(4248,440){\makebox(0,0){\strut{} 8}}%
      \put(264,1206){\rotatebox{90}{\makebox(0,0){\strut{}Force}}}%
      \put(2784,110){\makebox(0,0){\strut{}Particle number}}%
      \put(2784,1642){\makebox(0,0){\strut{}}}%
    }%
    \gplgaddtomacro\gplfronttext{%
    }%
    \gplbacktext
    \put(0,0){\includegraphics{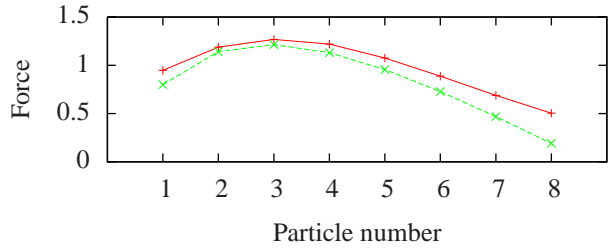}}%
    \gplfronttext
  \end{picture}%
\endgroup
}
\caption{\label{fig:fixed-spacing}Example of how the force  (arbitrary units) on particle $i$ in a chain of 8 particles due only to the right-going beam varies along the chain if the inter-particle spacing is constant. Mie scattering model (red, unbroken line) and our simple ansatz model (green, dotted line). Both plots show similar trends; neither is symmetric with respect to the center of the chain. This means that, when both beams are considered, there will \emph{not} be a net force of zero on a given particle, and so for this imposed uniform spacing the system will not be in equilibrium .}
\end{figure}

\begin{figure}
\centerline{
  \begingroup
  \setlength{\unitlength}{0.0500bp}%
  \begin{picture}(4320.00,1512.00)%
    \gdef\gplbacktext{}%
    \gdef\gplfronttext{}%
    \gplgaddtomacro\gplbacktext{%
      \csname LTb\endcsname%
      \put(770,264){\makebox(0,0)[r]{\strut{} 0.5}}%
      \put(770,1248){\makebox(0,0)[r]{\strut{} 2}}%
      \put(396,756){\rotatebox{90}{\makebox(0,0){\strut{}Spacing}}}%
      \put(2424,1138){\makebox(0,0){\strut{}}}%
    }%
    \gplgaddtomacro\gplfronttext{%
    }%
    \gplbacktext
    \put(0,0){\includegraphics{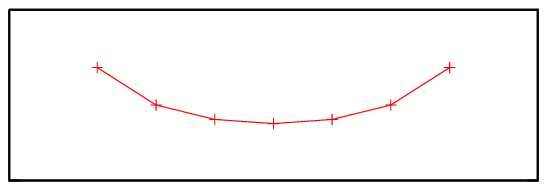}}%
    \gplfronttext
  \end{picture}%
\endgroup
}
\vspace{-0.5cm}
\centerline{
  \begingroup
  \setlength{\unitlength}{0.0500bp}%
  \begin{picture}(4320.00,2016.00)%
    \gdef\gplbacktext{}%
    \gdef\gplfronttext{}%
    \gplgaddtomacro\gplbacktext{%
      \csname LTb\endcsname%
      \put(770,660){\makebox(0,0)[r]{\strut{} 0.5}}%
      \put(770,1752){\makebox(0,0)[r]{\strut{} 1}}%
      \put(1240,440){\makebox(0,0){\strut{} 1}}%
      \put(1578,440){\makebox(0,0){\strut{} 2}}%
      \put(1917,440){\makebox(0,0){\strut{} 3}}%
      \put(2255,440){\makebox(0,0){\strut{} 4}}%
      \put(2593,440){\makebox(0,0){\strut{} 5}}%
      \put(2931,440){\makebox(0,0){\strut{} 6}}%
      \put(3270,440){\makebox(0,0){\strut{} 7}}%
      \put(3608,440){\makebox(0,0){\strut{} 8}}%
      \put(396,1206){\rotatebox{90}{\makebox(0,0){\strut{}Force}}}%
      \put(2424,110){\makebox(0,0){\strut{}Particle number}}%
      \put(2424,1642){\makebox(0,0){\strut{}}}%
    }%
    \gplgaddtomacro\gplfronttext{%
    }%
    \gplbacktext
    \put(0,0){\includegraphics{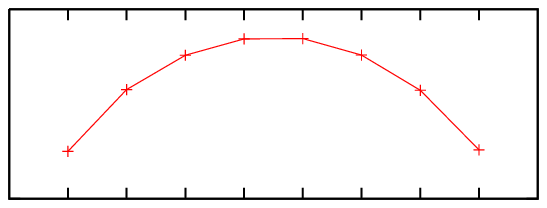}}%
    \gplfronttext
  \end{picture}%
\endgroup
}
\caption{\label{fig:variable-spacing}Example of how a force profile like those in Figure~\ref{fig:fixed-spacing} (arbitrary units) can be made symmetric by altering inter-particle spacings. Now, in contrast to Figure~\ref{fig:fixed-spacing}, when both beams are considered there will be a net force of zero on each particle, and so the system will be in equilibrium.}
\end{figure}

Having made these assumptions, we can test the predictions of this very simple model against the definitive calculations of a rigorous Mie scattering model, and against established experimental observations.
Figure~\ref{fig:fixed-spacing} illustrates how this model predicts the force to vary along a chain of particles with a constant inter-particle spacing, and an example of how the forces calculated from a Mie scattering model vary along a similar chain with constant particle spacing. It can be seen from this plot that requirement \ref{fig:symmetric-profile} (a symmetric force profile) is not satisfied in either model: the force pushing particle 1 \emph{inwards} is weaker than the force pushing particle $N$ \emph{outwards}, which means that the tendency will be for the outermost particles to move apart. The natural next step is therefore to allow the particle spacings to vary along the length of the chain, as they would do in real life in response to this repulsive force.
As Figure~\ref{fig:variable-spacing} shows, this approach allows a symmetric force profile to be produced, and leading to stable trapping of the chain with these slightly non-equilibrium spacings.

The first few particles in the chain act to focus the laser field onto the next particle in the chain, and hence initially the force rises sharply with particle index $i$. Particles towards the middle of the chain can be thought of as acting more like a (very inefficient) waveguide where the intensity is propagated from one particle to the next with some losses, which are compensated for by the re-focusing of additional background light. The intensity (and force) then drops again towards the end of the chain due to the increased particle spacings.

\begin{figure}
\centerline{
  \begingroup
  \setlength{\unitlength}{0.0500bp}%
  \begin{picture}(7200.00,3528.00)%
    \gdef\gplbacktext{}%
    \gdef\gplfronttext{}%
    \gplgaddtomacro\gplbacktext{%
      \csname LTb\endcsname%
      \put(726,660){\makebox(0,0)[r]{\strut{}-1}}%
      \put(726,1181){\makebox(0,0)[r]{\strut{} 0}}%
      \put(726,1702){\makebox(0,0)[r]{\strut{} 1}}%
      \put(726,2222){\makebox(0,0)[r]{\strut{} 2}}%
      \put(726,2743){\makebox(0,0)[r]{\strut{} 3}}%
      \put(726,3264){\makebox(0,0)[r]{\strut{} 4}}%
      \put(858,440){\makebox(0,0){\strut{} 0}}%
      \put(2386,440){\makebox(0,0){\strut{} 10}}%
      \put(3915,440){\makebox(0,0){\strut{} 20}}%
      \put(5443,440){\makebox(0,0){\strut{} 30}}%
      \csname LTb\endcsname%
      \put(220,1962){\rotatebox{90}{\makebox(0,0){\strut{}Force (fN)}}}%
      \put(3150,110){\makebox(0,0){\strut{}Inter-particle spacing ($\mu$m)}}%
      \put(3150,3154){\makebox(0,0){\strut{}}}%
      \put(1851,2899){\makebox(0,0)[l]{\strut{}2-particle stable spacing}}%
      \put(1928,1441){\makebox(0,0)[l]{\strut{}no 3-particle stable spacing}}%
    }%
    \gplgaddtomacro\gplfronttext{%
      \csname LTb\endcsname%
      \put(6213,3154){\makebox(0,0)[r]{\strut{}1 of 2}}%
      \csname LTb\endcsname%
      \put(6213,2934){\makebox(0,0)[r]{\strut{}2 of 2}}%
      \csname LTb\endcsname%
      \put(6213,2714){\makebox(0,0)[r]{\strut{}1 of 3}}%
      \csname LTb\endcsname%
      \put(6213,2494){\makebox(0,0)[r]{\strut{}3 of 3}}%
    }%
    \gplbacktext
    \put(0,0){\includegraphics{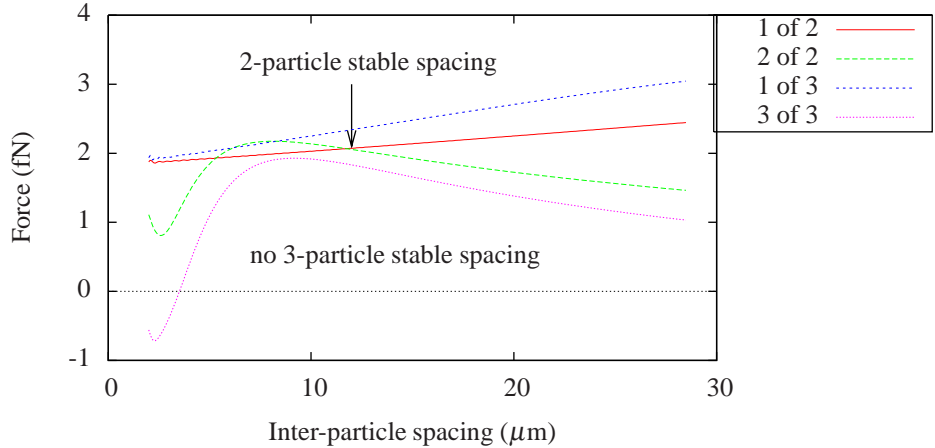}}%
    \gplfronttext
  \end{picture}%
\endgroup
}
\caption{\label{fig:collapse}Collapse of a longer chain, where a shorter chain would be stable. Parameters have been selected to give an extreme case where a two-particle chain is stable but a three-particle chain is not (parameters as~\cite{innsbruck}, but with 1.9~\um{} diameter spheres). The curves ``1 of 2" and ``2 of 2" show the forces due to the right-going beam for the particles in a two-particle chain. Since the force on the second particle is greater than the force on the first particle over a range of around 5-12~\um{} inter-particle spacing, there is a stable trapped configuration with an inter-particle spacing of about 12~\um{} (indicated with an arrow) when net effect of both beams is considered. The curves ``1 of 3" and ``3 of 3" show the same forces on the end particles of a three-particle chain. Since there is no spacing for which the force on the third particle is greater than the force on the first particle, the chain will collapse. Even though there is some enhancement of the force on the third particle over what it would be in the absence of the other particles, this is not enough to overcome the compressive force due to the effect of the beam on the first particle.}
\end{figure}
Finally, Figure~\ref{fig:collapse} shows how a chain above a critical length can collapse. The figure shows how the forces on the end particles in a short chain vary with inter-particle spacing for a particular set of experimental parameters (different to those used earlier, and carefully selected so the collapse occurs at an unusually short chain length). In this case a two-particle chain is supported, but if a third particle is added the chain will collapse until the spheres are in contact (an effect mentioned in~\cite{innsbruck}). As pointed out earlier, the compressive force on the first particle in the chain is greater if the chain has more particles in it (since that first particle is closer to the beam waist). We argued earlier that this would cause the inter-particle spacing to fall until the inter-particle repulsive force was increased enough to for the two forces to balance. However, for the particular parameters in this figure, at short inter-particle spacings there is in fact an \emph{attractive} force between neighboring particles, and so the three-particle chain collapses once the compressive forces have pushed the particles close enough to enter this regime. Our implicit assumption that the inter-particle light forces are repulsive (it was assumed that radiation pressure will dominate) has broken down; near-field gradient force effects have come into play, producing a net \emph{attractive} force between the spheres at close ranges. There is no repulsive force to support the chain, and it collapses.

\section{Conclusions}

We have explained the mechanisms behind the formation of optically bound particle chains in counter-propagating  Gaussian beam traps. The optical binding effect results from the balancing of repulsive effects from the light from one particle incident on the next particle in the chain and compressive effects due to the background trapping potential formed by the beams  (it could in fact be argued that this effect is not optical ``binding'' in the strict sense shown in early experiments~\cite{burns-binding-prl}, since the interaction here is largely a repulsive one, with stable chains only being formed due to the background harmonic potential of the trap). Here we have used a very simple model to successfully explain the trends of closer spacings as more particles are added to the chain, and of closer spacings in the center of a chain compared to near its edges.

While our simple model does not claim to agree precisely with experimental results and with the theoretical gold standard of Mie scattering calculations which we have also used (and has a number of parameters which must be tuned by hand), there is good qualitative agreement between them across a range of model parameters.
From this we can conclude that, while there is some influence from more sophisticated effects which can only be encapsulated in a full vector model based on rigorous solution of Maxwell's equations (such as Mie scattering), many of the properties of the trapped particle chains can be understood in terms of a simple scalar model.
This model can offer strong conceptual insights into the physical mechanisms which lead to the observed behavior, which had not previously been fully explained. As well as explaining how the inter-particle spacings are regulated, it explains the trend for closer spacings with larger $N$,
and the wider spacing close to either end of the chain.
It is \emph{only} through a simple model such as the one we have presented that the various complex effects in the experiment can be decoupled in order to understand \emph{why} optical binding occurs under these experimental conditions.

\section*{Acknowledgement}
This work was supported by the UK
Engineering and Physical Sciences Research Council.


\begin{thebibliography}{10}

\bibitem{burns-binding-science}
Michael~M. Burns, Jean-Marc Fournier, and Jene~A. Golovchenko.
\newblock Optical matter: Crystalization and binding in intense optical fields.
\newblock {\em Science}, 249:749--754, 1990.

\bibitem{Tatarkova}
S.~A. Tatarkova, A.~E. Carruthers, and K.~Dholakia.
\newblock One-dimensional optically bound arrays of microscopic particles.
\newblock {\em Phys. Rev. Lett.}, 89(28):283901, 2002.

\bibitem{Metzger:fluoro}
N.~K. Metzger, E.~M. Wright, W.~Sibbett, and K.~Dholakia.
\newblock Visualization of optical binding of microparticles using a
  femtosecond fiber optical trap.
\newblock {\em Optics Express}, 14(8):3677--3687, 2006.

\bibitem{Metzger:bistability2}
N.~K. Metzger, K.~Dholakia, and E.~M. Wright.
\newblock Observation of bistability and hysteresis in optical binding of two
  dielectric spheres.
\newblock {\em Phys. Rev. Lett.}, 96:068102, 2006.

\bibitem{Mellor:polarization}
Christopher~D. Mellor, Thomas~A. Fennerty, and Colin~D. Bain.
\newblock Polarization effects in optically bound particle arrays.
\newblock {\em Opt. Express}, 14:10079--10088, 2006.

\bibitem{ng:clusters}
Jack Ng, Z.~F. Lin, C.~T. Chan, and Ping Sheng.
\newblock Photonic clusters formed by dielectric microspheres: Numerical
  simulations.
\newblock {\em Phys. Rev. B}, 72:085130, 2005.

\bibitem{cp-dynamics}
M.~Kawano, J.~T. Blakely, R.~Gordon, and D.~Sinton.
\newblock Theory of dielectric micro-sphere dynamics in a dual-beam optical
  trap.
\newblock {\em Opt. Express}, 16:9306--9317, 2008.

\bibitem{us:evanescent}
J.~M. Taylor, L.~Y. Wong, C.~D. Bain, and G.~D. Love.
\newblock Emergent properties in optically bound matter.
\newblock {\em Opt. Express}, 16:6921--6929, 2008.

\bibitem{bessel-binding}
V.~Kar\'asek, O~Brzobohat\'y, and P.~Zem\'anek.
\newblock Longitudinal optical binding of several spherical particles studied
  by the coupled dipole method.
\newblock {\em J. Opt. A}, 11:034009, 2009.

\bibitem{nanoparticle-review}
Maria Dienerowitz, Michael Mazilu, and Kishan Dholakia.
\newblock Optical manipulation of nanoparticles: A review.
\newblock {\em J. Nanophotonics}, 2:021875, 2008.

\bibitem{burns-binding-prl}
Michael~M. Burns, Jean-Marc Fournier, and Jene~A. Golovchenko.
\newblock Optical binding.
\newblock {\em Phys. Rev. Lett.}, 63(12):1233--1236, 1989.

\bibitem{McGloin:1d-binding}
D.~McGloin, A.~E. Carruthers, K.~Dholakia, and E.~M. Wright.
\newblock Optically bound microscopic particles in one dimension.
\newblock {\em Phys. Rev. E}, 69:021403, 2004.

\bibitem{cizmar:bessel}
Tom\'as {\v C}i{\v z}m\'ar, V{\v e}ra Koll\'arov\'a, Zden{\v e}k Bouchal, and
  Pavel Zem\'anek.
\newblock Sub-micron particle organization by self-imaging of non-diffracting
  beams.
\newblock {\em New J. Phys.}, 8:43, 2006.

\bibitem{mackowski:translation}
Daniel~W. Mackowski.
\newblock Analysis of radiative scattering for multiple sphere configurations.
\newblock {\em Proc. R. Soc. London, Ser. A}, 433:599--614, 1991.

\bibitem{barton:force}
J.~P. Barton, D.~R. Alexander, and S.~A. Schaub.
\newblock Theoretical determination of net radiation force and torque for a
  spherical particle illuminated by a focused laser beam.
\newblock {\em J. Appl. Phys.}, 66(10):4594--4602, 1989.

\bibitem{innsbruck}
W.~Singer, M.~Frick, S.~Bernet, and M.~Ritsch-Marte.
\newblock Self-organized array of regularly spaced microbeads in a
  fiber-optical trap.
\newblock {\em J. Opt. Soc. Am. B}, 20(7):1568--1574, 2003.

\end{thebibliography}
\end{document}